\documentclass[a4paper,10pt]{article}

\usepackage{amssymb, amsthm}
\usepackage{bbm}
\usepackage{cite}
\usepackage{latexsym}
\usepackage{amsmath}
\usepackage{hyperref}
\usepackage{graphicx}
\usepackage{xspace}
\usepackage[usenames,dvipsnames]{xcolor}
\usepackage{mathtools}
\usepackage{ifthen}
\usepackage[utf8]{inputenc}
\usepackage[english]{babel}
\usepackage[labelfont={bf}]{caption}
\usepackage{subcaption}
\usepackage{titling}

\addto\captionsngerman{}

\numberwithin{equation}{section}
\numberwithin{figure}{section}
\numberwithin{table}{section}

\newcommand{\adsds}[1]{\ensuremath{\text{(A)dS}_{#1}}}
\newcommand{\np}{nearly parallel\xspace}
\newcommand{\Np}{Nearly parallel\xspace}
\newcommand{\gstr}{$G_2$-structure\xspace}
\newcommand{\gmf}{$G_2$-manifold\xspace}
\newcommand{\gmfs}{$G_2$-manifolds\xspace}
\newcommand{\ghol}{$G_2$-holonomy\xspace}

\newcommand{\hq}{\ensuremath{\widehat{Q}}}
\newcommand{\tr}{\ensuremath{\text{tr}}}
\newcommand{\unity}{\mathbbm{1}}
\newcommand{\RR}{\ensuremath{\mathbb{R}}}

\newcommand{\ub}[1]{_{(#1)}}

\newcommand{\und}{\qquad\text{and}\qquad}
\newcommand{\oder}{\qquad\text{or}\qquad}
\newcommand{\for}{\qquad\text{for}\quad}
\newcommand{\with}{\quad\text{with}\quad}
\newcommand{\ap}{\ensuremath{{\alpha^\prime}}\xspace}
\newcommand{\vol}[1]{\ensuremath{\text{vol}^{(#1)}}}
\newcommand{\dd}{\text{d}}
\newcommand{\g}[2]{\ensuremath{{\langle #1 , #2 \rangle}}}

\DeclareMathOperator{\Scal}{Scal}
\DeclareMathOperator{\Ric}{Ric}
\DeclareMathOperator{\diag}{diag}

\title{Heterotic $G_2$-manifold compactifications with fluxes and fermionic condensates}
\author{Karl-Philip Gemmer and Olaf Lechtenfeld}
\date{\today}

\begin{document}

\begin{titlepage}
\setcounter{page}{0}
\begin{flushright}
ITP--UH--14/13\\
\end{flushright}

\vskip 2.0cm

\begin{center}

{\huge\bf Heterotic $G_2$-manifold compactifications\\[.5cm] with fluxes and fermionic condensates}

\vspace{12mm}

{\LARGE 
Karl-Philip Gemmer${}^\dagger$\ and \ 
Olaf~Lechtenfeld${}^{\dagger\times}$
}
\\[8mm]
\noindent ${}^\dagger${\em
Institut f\"ur Theoretische Physik\\
Leibniz Universit\"at Hannover \\
Appelstra\ss{}e 2, 30167 Hannover, Germany }
\\[8mm]
${}^\times${\em
Riemann Center for Geometry and Physics\\
Leibniz Universit\"at Hannover \\
Appelstraße 2, 30167 Hannover, Germany }

\vspace{12mm}

\begin{abstract}
\noindent 
We consider flux compactifications of heterotic string theory in the presence of fermionic condensates on $M_{1,2} \times X_7$ with both factors carrying a Killing spinor. In other words, $M_{1,2}$ is either de Sitter, anti-de Sitter or Minkowski, and $X_7$ possesses a \np \gstr or has \ghol.   We solve the complete set of field equations and the Bianchi identity to order \ap.  The latter is satisfied via a non-standard embedding by choosing the gauge field to be a $G_2$-instanton.  It is shown that none of the solutions to the field equations is supersymmetric.
\end{abstract}

\end{center}
\end{titlepage}

\thispagestyle{empty}
\newpage

\section{Introduction and summary}

Many compactifications of string theory suffer from the severe problem of moduli stabilization, the existence of scalar fields whose vacuum expectation values are not fixed by a potential.  A promising method for the heterotic string to stabilize these scalar fields is the introduction of fluxes and fermionic condensates, i.e.~vacuum expectation values of some tensor fields and some fermionic bilinears along the internal manifold.  Without fluxes and condensates, the Killing spinor equations demand the internal manifold to have reduced holonomy, e.g.~SU(3) ($G_2$) for compactifications on a six-(seven-)dimensional manifold.  Fluxes lead to a deformation of the internal manifold,  resulting in an internal space with only reduced structure group but no reduced holonomy.  

For compactifications on six-dimensional manifolds, deformations to non-K\"ahler SU(3)-manifolds have already been studied in detail \cite{Strominger1986,Cardoso2003}.  Also the effect of implementing gaugino and dilatino condensates has been analyzed \cite{Cardoso2004,Frey2005,Manousselis2006,Lechtenfeld2010,Chatzistavrakidis2012}.  Here, we investigate which aspects of these results carry over to compactifications on seven-dimensional manifolds~$X_7$.  

More specifically, we discuss compactifications on manifolds with \ghol as well as on their deformations to \np \gmfs, in the presence of fermionic condensates.  Assuming the space-time background to be a product of $X_7$ and a maximally symmetric Lorentzian space $M_{1,2}$, we solve the field equations to order~\ap and discuss the conditions under which the solutions preserve supersymmetry.  The Bianchi identity is also satisfied to guarantee the absence of anomalies.  The gauge field is taken to be a generalized instanton on the internal manifold $X_7$.  This choice allows us to solve the Bianchi identity by a non-standard embedding and immediately takes care of the Yang-Mills equation.  Furthermore, it also ensures the vanishing of the gaugino supersymmetry variation.

There are several aspects in which the considered compactifications to three dimensions differ from those to four dimensions.  Most importantly, the fermionic condensates cannot be restricted to the internal manifold $X_7$ but must extend to $M_{1,2}$.  Therefore, the field equations do not decouple into separate equations on $M_{1,2}$ and $X_7$.  As a first consequence, the equations of motion allow not only for anti-de Sitter solutions but admit de Sitter and Minkowski space-times as well.  Secondly, the radius of the de Sitter or anti-de Sitter space is not fixed but related to the amplitudes of the condensates and $H$-flux by the equations of motion.  It turns out that none of these heterotic vacua is supersymmetric.

The paper is organized as follows.  In Section~\ref{sec:hetStr+femCond}, we briefly review the action and the equations of motion of heterotic supergravity to first order in~\ap.  We decompose the fields and their equations according to the space-time factorization $M_{1,2} \times X_7$, including the effect of gaugino and dilatino condensates.  The geometric properties of manifolds with \gstr are the subject of Section~\ref{sec:g2manifolds}.  In Section~\ref{sec:hetStr_on_adsxX7} we solve the heterotic equations of motion for the six possible combinations of $M_{1,2}$ being either de Sitter, anti-de Sitter or Minkowski space-times and for $X_7$ carrying either \ghol or just a \np $G_2$-structure.  Furthermore, we present the conditions for supersymmetric solutions and compute the fermion masses for all considered backgrounds.

\section{Heterotic string with fermionic condensates}
\label{sec:hetStr+femCond}

\paragraph{Action and equations of motion.}
The low-energy field theory limit of heterotic string theory is given by $d=10$, $\mathcal N =1$ supergravity coupled to a super-Yang-Mills multiplet, and it is defined on a ten-dimensional space-time $M$.  The supergravity multiplet consists of the graviton $g$, which is a metric on $M$, the left-handed Rarita-Schwinger gravitino $\Psi$, the Kalb-Ramond two-form field $B$, the scalar dilaton $\phi$ and the right-handed Majorana-Weyl dilatino $\lambda$.  Moreover, the vector supermultiplet consists of the gauge field one-form $A$ and its superpartner, the left-handed Majorana-Weyl gaugino $\chi$.

Rather than presenting the full action describing the propagation and interactions of the above fields \cite{Bergshoeff1982,Chapline1983}, we shall restrict ourselves to the part which is relevant for our purposes.  In this paper we shall consider vacuum solutions where the fermionic expectation values are forced to vanish by requiring Lorentz invariance, but certain fermionic bilinears may acquire non-trivial vacuum expectation values.  However, these vacuum expectation values will not involve the gravitino, whence we set the gravitino to zero from the very beginning, $\Psi = 0$.  Then, in the string frame, the low-energy action up to and including terms of order \ap reads as \cite{Bergshoeff1989}
\begin{multline}\label{eq:hetaction}
\mathcal S=  \int_M\!\!\dd^{10}x \sqrt{\det g}\ e^{-2\phi} 
   \Big[  \text{Scal} + 4|\dd\phi|^2 - \tfrac 12|H|^2 + \tfrac12 (H,\Sigma) - 2 (H,\Delta)+ \tfrac 14 (\Sigma,\Delta) - \tfrac 18|\Sigma|^2 +  \\ 
       + \tfrac14\ap\left(\tr  |\tilde R|^2 -\tr\left(|F|^2 -2 \langle\chi, \mathcal D \chi\rangle - \tfrac 13 \langle\lambda,\gamma^M\gamma^{AB}F_{AB}\gamma_M\lambda\rangle\right)\right)
       +8\langle\lambda,\mathcal D \lambda\rangle \Big]  \:.
\end{multline}
Here, Scal is the scalar curvature of the Levi-Civita connection $\Gamma^g$ on $TM$.  Furthermore, $\tilde R$ is the curvature form of a connection $\tilde\Gamma$ on the $TM$.  The choice of this connection is ambiguous and will be discussed in Section \ref{sec:hetStr_on_adsxX7}.  Here and in the following, traces are taken in the adjoint representation of SO(9,1) or of the gauge group, respectively, depending on the context. 

The field strength $H$ is defined by
\begin{align}
\label{eq:H=dB+...}
  H = \dd B + \tfrac 14 \ap \left( \omega_{CS}(\tilde \Gamma) - \omega_{CS}(A) \right) \:,
\end{align}
where the Chern-Simons forms of the connections $\tilde \Gamma$ and $A$ are given by
\begin{align}
  \omega_{CS}(\tilde \Gamma) = \tr\left(\tilde R \wedge \tilde \Gamma -\tfrac 23 \tilde \Gamma \wedge \tilde \Gamma \wedge \tilde \Gamma \right) \quad\text{and}\quad
  \omega_{CS}(A)            = \tr\left( F \wedge  A -\tfrac 23 A \wedge A \wedge A \right) \:.
\end{align}
For any two $p$-forms $\alpha$, $\beta$ we use the definitions
\begin{equation}
  (\alpha,\beta) := \frac1{p!} \alpha_{M_1,\ldots,M_p} \beta^{M_1,\ldots,M_p} \:,\quad |\alpha|^2 := (\alpha,\alpha) \:.
\end{equation}
$\mathcal D = \gamma^M\nabla_M$ denotes the Dirac operator coupled to $\Gamma^g$ and $A$.  Finally, we have defined the fermion bilinears
\begin{align}
 \Sigma = \tfrac1{24}\ap \tr \langle \chi, \gamma_{MNP} \chi \rangle \, e^{MNP} \und 
 \Delta = \tfrac16 \langle \lambda, \gamma_{MNP} \lambda \rangle e^{MNP} \:,
\end{align}
with $e^{MNP}\equiv e^M \wedge e^N \wedge e^P$ and $\{ e^M \}$ being an othonormal frame on the space-time background $M$.  By $\langle \cdot , \cdot \rangle$ we denote the inner product of spinors.

\begin{subequations}
\label{eq:SUSYconditions}
The action \eqref{eq:hetaction} is invariant under $\mathcal N=1$ supersymmetry transformations, which act on the fermions as
\begin{align}
  \label{eq:gravitinoVariation}
  \delta\Psi_M &= \nabla_M\epsilon - \tfrac18 H_{MNP}\gamma^{NP} \epsilon + \tfrac1{96} \Sigma \cdot \gamma_M\epsilon \:, \\
  \label{eq:dilatinoVariation}
  \delta\lambda &= -\tfrac{\sqrt2}4(d\phi-\tfrac1{12}H- \tfrac1{48}\Sigma+\tfrac1{48}\Delta) \cdot \epsilon \:, \\
  \label{eq:gauginoVariation}
  \delta\chi &= -\tfrac14F \cdot \epsilon + \g\chi\lambda \, \epsilon - \g\epsilon\lambda\,\chi + \g\chi{\gamma_M\epsilon} \, \gamma^M\lambda \:,
\end{align}
where $\epsilon$ is the supersymmetry parameter, a left-handed Majorana-Weyl spinor.
\end{subequations}

The equations of motion may be obtained by varying the action \eqref{eq:hetaction} and take the form
\begin{subequations}
\label{eq:10D-eom}
\begin{align}
&\begin{multlined}[c][.8\textwidth]
\label{eq:einstein eq}
 0 = \text{Ric}_{MN}  +2 (\nabla \dd\phi)_{MN} -\tfrac 1{8} (H-\tfrac 12\Sigma+2\Delta)_{PQ(M}{H_{N)}}^{PQ}+ \\
\tfrac14\ap \Big[ \tilde R_{MPQR}{\tilde R_N}^{\ PQR} - \text{tr}\big(F_{MP} {F_N}^P +\tfrac 12\g\chi{\gamma_{(M} \nabla_{N)}\chi}\big)\Big]+2\g\lambda{\gamma_{(M}\nabla_{N)}\lambda} \:,
\end{multlined}\\
&\begin{multlined}[c][.8\textwidth]
\label{eq:dilaton eq}
 0 = \text{Scal}  - 4 \Delta \phi  +4|\dd\phi|^2-\tfrac 12|H|^2+\tfrac{1}{2}(H,\Sigma)-2(H,\Delta)+\tfrac 14(\Sigma,\Delta)-\tfrac{1}{8}|\Sigma|^2 \\
  +\tfrac14\ap\text{tr}\Big[ |\tilde R|^2 - |F|^2 - 2\g\chi{\mathcal D\chi}\Big]+8\g\lambda{\mathcal D\lambda} \:, 
\end{multlined}\\
\label{eq:YM eq}
&\begin{multlined}[c][.8\textwidth]
 0 =  e^{2\phi}\dd\ast (e^{-2\phi}F) + A\wedge\ast F- \ast F\wedge A+\ast (H-\tfrac{1}{2}\Sigma+2\Delta)\wedge F  \:,
\end{multlined}\\
&\begin{multlined}[c][.8\textwidth]
 0 = \dd\ast e^{-2\phi} (H-\tfrac{1}{2}\Sigma+2\Delta) \:,
\end{multlined} \\
&\begin{multlined}[c][.8\textwidth]
 \label{eq:dirac eq gaugino}
 0 = \big( \mathcal D - \tfrac 1{24} (H-\tfrac 12\Sigma+\tfrac 12\Delta)\cdot \big) \: e^{-2\phi}\chi \:,
\end{multlined} \\
&\begin{multlined}[c][.8\textwidth]
  \label{eq:dirac eq dilatino}
 0 = \big( \mathcal D - \tfrac 1{24} (H-\tfrac 18\Sigma) \cdot \big) \:e^{-2\phi}\lambda \:.
\end{multlined}
\end{align}
\end{subequations}
They are complemented by the Bianchi identity 
\begin{align}
   \label{eq:BianchiEQ}
   \dd H = \tfrac14 \ap\left( \tr(\tilde R \wedge \tilde R) - \tr(F \wedge F) \right)\:,
\end{align}
which follows from the definition \eqref{eq:H=dB+...}.  In order to simplify the equations of motion, we will assume for the remainder of the paper that the dilation vanishes, i.e.
\begin{equation}
  \phi=0 \:.
\end{equation}

\paragraph{Space-time and spinor factorization.} We will consider space-time backgrounds $M$ of the form
\begin{equation}
  M = M_{1,2} \times X_7 \:,
\end{equation}
with $M_{1,2}$ being a maximally symmetric Lorentzian manifold and $X_7$ a being seven-dimensional compact Riemannian manifold.  Furthermore, we assume that $X_7$ possesses a nowhere vanishing real Killing spinor\footnote{Note that the notion of real an imaginary Killing spinors differs from parts of the mathematical literature, due to a different sign in the definition of the Clifford algebra (see \eqref{eq:def clifford algebra}).}, i.e.~a spinor satisfying
\begin{equation}
 \nabla^g_X \eta = i\,\mu_2 \,  X\cdot \eta 
\end{equation}
for some real constant $\mu_2$.  If $\mu_2=0$, then $X_7$ admits \ghol.  Manifolds with non-vanishing $\mu_2$ are known as \np \gmfs (see Section \ref{sec:g2manifolds}).  We denote the metric on $M$ by $g$, whereas the metrics on $M_{1,2}$ and $X_7$ will be labeled by $g_3$ and $g_7$, respectively.

The factorization of the space-time background $M$ induces a splitting of the SO(9,1) Clifford algebra.  We employ a standard representation of the SO(9,1) Clifford algebra 
\begin{equation}
\label{eq:def clifford algebra}
  \{\gamma^A,\gamma^B\} = 2 \eta^{AB} = 2 \diag(-1,+1,\ldots,+1)^{AB}
\end{equation}
via
\begin{subequations}
\begin{align}
 \gamma^\mu &= \gamma\ub3^\mu \otimes \unity_8 \otimes \sigma_2 \for \mu=0,1,2 \:,\\
 \gamma^{a+2} &= \unity_2 \otimes \gamma\ub7^a \otimes \sigma_1 \for a=1,\ldots,7 \:.
\end{align}
\end{subequations}
Here, $\unity_n$ are $n\times n$-unit matrices and $\gamma\ub7^a$ are SO(7) gamma matrices.   Furthermore, $\gamma\ub3^\mu$ and $\sigma_i$ denote the SO(2,1) gamma matrices and Pauli matrices, respectively:
\begin{alignat}{3}
 \gamma\ub3^0 &= \begin{pmatrix} 0 & -1 \\  1 &  0 \end{pmatrix} \:,\qquad &
 \gamma\ub3^1 &= \begin{pmatrix} 0 &  1 \\  1 &  0 \end{pmatrix} \:,\qquad &
 \gamma\ub3^2 &= \begin{pmatrix} 1 &  0 \\  0 & -1 \end{pmatrix} \:, \\
 \sigma^1 &= \begin{pmatrix} 0 &  1 \\  1 &  0 \end{pmatrix} \:,\qquad &
 \sigma^2 &= \begin{pmatrix} 0 & -i \\  i &  0 \end{pmatrix} \:,\qquad &
 \sigma^3 &= \begin{pmatrix} 1 &  0 \\  0 & -1 \end{pmatrix} \:.
\end{alignat}
The chirality and the charge conjugation operator in this representation read
\begin{equation}
 \Gamma^{11} = \unity_2 \otimes \unity_8 \otimes \sigma^3 \und
 \mathcal C =  C\ub3 \otimes \mathcal C\ub7 \otimes \mathcal \unity_2\:,
\end{equation}
with
\begin{equation}
 \mathcal C\ub7 = i \gamma\ub7^2\gamma\ub7^4\gamma\ub7^6 \und
 \mathcal C\ub3 = \gamma\ub3^0 \:.
\end{equation}

We assume that the dilatino and gaugino decompose as
\begin{subequations}
\label{eq:decomposition_of_fermions}
\begin{align}
 \chi    &= \widehat\chi    \otimes \eta \otimes (1,0)^t \:, \\
 \lambda &= \widehat\lambda \otimes \eta \otimes (0,1)^t \:,
\end{align}
\end{subequations}
with $\widehat\chi$ and $\widehat\lambda$ being Grassmann-valued SO(2,1) Majorana spinors and $\eta$ being the real Majorana Killing spinor on $X_7$.  The last factor in the products accounts for the opposite chirality of the gaugino and the dilatino.  The only non-vanishing spinor bilinears constructed with a single spinor $\eta$ are $\langle\eta,\eta\rangle$ and $\langle\eta,\gamma_{abc}\eta\rangle$.   Hence, $\Sigma$ and $\Delta$ simplify to
\begin{align}
\label{eq:gaugino condensate m}
 \Sigma &= m \: ( - \vol3 + \: Q  ) \:, \\
 \label{eq:dilatino condensate n}
 \Delta &= n \: ( - \vol3 + \: Q  ) \:,
\end{align}
with
\begin{equation}
 m = \tfrac1{24}\ap \tr\langle\widehat\chi,\widehat\chi\rangle \und 
 n = \tfrac16 \langle\widehat\lambda,\widehat\lambda\rangle \:.
\end{equation}
Here, $\vol3$ is the volume form on $M_{1,2}$, and 
\begin{equation}
  Q = -\frac{i}{3!} \langle \eta , (\gamma\ub7)_{mnp}\, \eta \rangle \, e^{mnp} \:.
\end{equation}
The properties of the three-form $Q$ will be discussed in Section \ref{sec:g2manifolds}.

From now on, we will replace all terms depending on fermion bilinears by their quantum expectation values.  Furthermore, we assume that the only non-vanishing expectation values are $\Sigma$ and $\Delta$.  Note that the form of the condensates \eqref{eq:gaugino condensate m} and \eqref{eq:dilatino condensate n} differs crucially from condensates considered previously in compactifications to four-dimensional space-times.  In the latter case one may consistently confine the condensate to the compactification space.  For a compactification to a three-dimensional space-time, however, a non-vanishing condensate must always have a space-time component, due to the fact that
\begin{equation}
\label{eq:clifford_action_of_vol3}
 \Gamma^0\Gamma^1\Gamma^2 = - \unity \:.
\end{equation}

\paragraph{Geometric data of maximally symmetric Lorentzian manifolds.}
For future reference we review some aspects of the geometry of maximally symmetric Lorentzian manifolds, i.e.~de Sitter, anti-de Sitter and Minkowski spaces.  These spaces possess a Killing spinor with Killing number $\mu_1$, meaning a spinor $\zeta$ satisfying
\begin{align}
  \nabla^g_X \zeta = \mu_1 X \cdot \zeta \:.
\end{align}
For de Sitter space, $\mu_1$ is real, whereas for anti-de Sitter space, $\mu_1$ is purely imaginary.  It vanishes on Minkowski space.  We define the (anti-)de Sitter radius $|\rho_1|$ by setting
\begin{equation}
  \mu_1 =
 \begin{cases}
      \rho_1^{-1} & \text{for de Sitter space } \\
   i\:\rho_1^{-1} & \text{for anti-de Sitter space}
 \end{cases} \:.
\end{equation}
The Ricci tensor and scalar curvature of (anti-)de Sitter space are given by
\begin{equation}
  \Scal_3 = \pm\frac{24}{\rho_1^2}  \und
  \Ric_3 = \pm\frac{8}{\rho_1^2} \: g_3 \:.
\end{equation}
Moreover, the curvature dependent quantities entering the Einstein and dilaton equations \eqref{eq:einstein eq} and \eqref{eq:dilaton eq} are given by 
\begin{equation}
   (R_3)_{\mu\alpha\beta\gamma}  (R_3)_\nu{}^{\alpha\beta\gamma} = \frac{64}{\rho_1^4} \: g_3 \und
   |R|^2 = \frac{192}{\rho_1^4} \:.
\end{equation}
Note that the curvature only depends on even powers of $\rho_1$ and, hence, the sign of $\rho_1$ will only enter the supersymmetry variations but not the equations of motion.

\section{The geometry of manifolds with \gstr}
\label{sec:g2manifolds}

\paragraph{\gmfs.} Manifolds with a \gstr are by definition seven-dimensional Riemannian manifolds possessing a nowhere vanishing $G_2$-invariant three-form $Q$.  Equivalently, they can be defined by the existence of a nowhere vanishing spinor $\eta$.  One can always find an orthonormal frame $\{e^a\}$ such that the three-form $Q$ can be written as
\begin{equation}
  \label{eq:psi-std-form}
  Q = e^{135} + e^{425} + e^{416} + e^{326} + e^{127} + e^{347} + e^{567} \:.
\end{equation}
Here and in the following we use the abbreviation $e^{abc} \equiv e^a \wedge e^b \wedge e^c$.  The \gstr defined by $Q$ is compatible with the metric $g_7$ in the sense that 
$*((X\lrcorner Q)\wedge(Y\lrcorner Q)\wedge Q) = 6\, g_7(X,Y)$ for all vector fields $X$ and $Y$.  The three-form $Q$ is related to the spinor $\eta$ by
\begin{equation}
  Q = -\frac{i}{3!} \langle \eta , \gamma_{mnp} \eta \rangle \, e^{mnp} \:.
\end{equation}
{}From this, one can deduce the action of $Q$ on $\eta$ under Clifford multiplication:
\begin{align}
\label{eq:Psi.eta}
 Q \cdot \eta &= 7i \;\eta \:,\\
 Q \cdot (X \cdot \eta) &= -i\; X \cdot \eta \:, \\
\label{eq:X_Psi.eta}
 (X \lrcorner Q) \cdot \eta &= 6i \; X \cdot \eta \:.
\end{align}

\paragraph{Form decomposition.}
Under the action of $G_2$ the spaces of $p$-forms on $X_7$, $\Lambda^p$, split into irreducible representations.  For the subsequent discussion we need the decompositions \cite{Fernandez1982}
\begin{align}
 \Lambda^2 &= \Lambda_{(7)}^2 \oplus \Lambda_{(14)}^2 \:,\\
 \Lambda^4 &= \Lambda_{(1)}^4 \oplus \Lambda_{(7)}^4 \oplus \Lambda_{(27)}^4 \:,
\end{align}
with
\begin{subequations}
\begin{align}
  \Lambda_{(7)}^2 &= \{v \lrcorner Q | v\in T_mX_7 \} \:,\\
  \Lambda_{(14)}^2 &= \{\beta \in \Lambda^2 | (*Q)\wedge\beta = 0 \}
\end{align}
\end{subequations}
and
\begin{subequations}
\begin{align}
 \Lambda_{(1)}^4 &= \{\mu *Q | \mu\in\RR \} \:,\\
 \Lambda_{(7)}^4 &= \{\alpha\wedge Q | \alpha\in T^*_mX_7 \} \:,\\
 \Lambda_{(27)}^4 &= \{ \gamma \in \Lambda^4 | \gamma\wedge Q=0 \text{ and } *\gamma\wedge Q =0 \} \cong S_0^2 \:.
\end{align}
\end{subequations}
The subscripts of $\Lambda^p$ label the dimension of the $G_2$ representation, and $S_0^2$ is the space of traceless symmetric two-tensors.

\paragraph{Manifolds with \ghol.}
\gstr manifolds are classified by the derivative of the three-form $Q$ or equivalently of the spinor $\eta$.  On manifolds with \ghol, the three-form $Q$ is closed and coclosed, and the spinor $\eta$ is parallel everywhere with respect to the Levi-Civita connection, 
\begin{equation}
  \label{eq:g2 holonomy <-> dQ=0}
  \dd Q = \dd * Q = 0 \und
  \nabla^g\eta = 0 \:.
\end{equation}
On the other hand, either the closedness and coclosedness of $Q$ or the vanishing of $\nabla^g\eta$ imply that the holonomy is contained in $G_2$.  As a result of \eqref{eq:g2 holonomy <-> dQ=0}, manifolds with \ghol are Ricci flat.

\paragraph{\Np \gmfs.}
By definition, the exterior derivative of the three-form $Q$ on a \np \gmf \cite{Friedrich1997} is proportional to its Hodge dual, 
\begin{equation}
  \label{eq:dPsi=*Psi}
  \dd Q = -8 \mu_2 * Q 
\end{equation}
for some $\mu_2\in\RR\backslash \{0\}$.  This is equivalent to demanding the spinor $\eta$ to be a real Killing spinor with nonzero Killing number $\mu_2$, 
\begin{equation}
  \label{eq:defKillingSpinor}
  \nabla^g_X \eta =  i\,\mu_2 \,  X\cdot \eta \:.
\end{equation}
Since the Killing number transforms under a conformal transformation $g\rightarrow \lambda^2 g$ as $\mu_2 \rightarrow \lambda^{-1} \mu_2$, its inverse modulus $|\mu_2^{-1}|$ measures the size of the \np \gmf.  Thus, analogously to the parameter $\rho_1$ for \adsds3-spaces, on \np \gmfs we set 
\begin{equation}
\mu_2=\rho_2^{-1} \:.
\end{equation}

On every \np \gmf we have two prominent connections: the Levi-Civita connection $\nabla^g$ and the canonical connection $\nabla^C$.  The latter is defined as the unique metric connection with respect to which $Q$ and equivalently $\eta$ are constant \cite{Friedrich1997}.  It differs from the Levi-Civita connection by a totally skew-symmetric torsion $T$,
\begin{equation}
  \label{eq:cannonical-connection_vs_LC-connection}
  g_7(\nabla^C_XY,Z) = g_7(\nabla^g_XY,Z) + T(X,Y,Z) \with  T = -\tfrac43\, \mu_2 \: Q  \:. 
\end{equation}
For later reference, we additionally define an interpolating connection
\begin{equation}
  \nabla^\kappa = \kappa \nabla^g + (1-\kappa) \nabla^C 
  \qquad\textrm{with}\quad \kappa\in\RR\:.
\end{equation}

\paragraph{Curvature of \np \gmfs.} With respect to the Levi-Civita connection, \np \gmfs are Einstein \cite{Friedrich1997} with Einstein constant $24 \mu_2^2$, 
\begin{equation}
  \Ric^g = 24 \mu_2^2\, g_7 \:.
\end{equation}
Note that, analogously to de Sitter and anti-de Sitter spaces, the curvature only depends on the square of $\mu_2$ and, hence, the sign of $\mu_2$ or equivalently $\rho_2$ will enter only the supersymmetry variations but not the equations of motion.

The canonical connection $\nabla^C$ is a $G_2$-instanton connection, meaning that its curvature form $R^C$ satisfies
\begin{equation}
\label{eq:G2-instanton eq}
  * R^C = -Q\wedge R^C \:.
\end{equation}
This is equivalent to $R^C$ annihilating the Killing spinor $\eta$ by its Clifford action, 
\begin{equation}
  R^C \cdot \eta = 0 \:.
\end{equation}

Beyond that, we also need the quantities $R_{abcd}{R_e}^{bcd}$, $|R|^2$ and $\tr(R\wedge R)$ to discuss the field equations \eqref{eq:10D-eom} of the action \eqref{eq:hetaction}.  We do not calculate these quantities explicitly, but relate them for the connection $\nabla^\kappa$ to their value for the canonical connection.  In this way, the curvature tensor $R^\kappa$ of the connection $\nabla^\kappa$ reads
\begin{equation}
 \label{eq:curvature_tildeR} 
 (R^\kappa)_{abcd} = (R^C)_{abcd} -  \frac{16}{9}\,\mu_2^2\,\kappa^2  \, (*Q)_{abcd} + \frac49\,\mu_2^2\,\kappa^2 \, ({(g_7)}_{ac}{(g_7)}_{bd} - {(g_7)}_{ad}{(g_7)}_{bc}) \:.
\end{equation}
For $\kappa=1$ we obtain the curvature of the Levi-Civita connection.  Using the identities for $Q$ in \eqref{eq:psi-identities} we obtain
\begin{align}
 (R^\kappa)_{acde}{(R^\kappa)_b}^{cde} &= (R^C)_{acde} {(R^C)_b}^{cde} - \frac{64}{9}\,\mu_2^4\,\kappa^2\, (16-11\kappa^2) \: {(g_7)}_{ab} \:,\\[.2cm]
 |R^\kappa|^2 &= |R^C|^2 - \frac{448}9\,\mu_2^4\, \kappa^2\,(16-11\kappa^2)  \:,\\[.2cm]
 \tr(R^\kappa\wedge R^\kappa) &= \tr(R^C \wedge R^C) - \frac{256}{27}\, \mu_2^4\,\kappa^2\, (1+\kappa^2) \: (*Q) \:.
\end{align}
At this point we also note the following fact\cite{Bryant2006}: For a generic two-form $\beta \in \Lambda_{(14)}^2$, the wedge product $\beta\wedge\beta$ lies in $\Lambda_{(1)}^4\oplus\Lambda_{(27)}^4$.  Furthermore, the components of $\beta\wedge\beta$ in $\Lambda_{(1)}^4$ and $\Lambda_{(27)}^4$ cannot vanish separately.  As the curvature form of the canonical connection is in $\Lambda_{(14)}^2 \otimes \mathfrak g_2$, this also applies to $\tr(R^C \wedge R^C)$.  Moreover, as $\tr(R^\kappa\wedge R^\kappa)$ differs from $\tr(R^C\wedge R^C)$ by a term in $\Lambda_{(1)}^4$, the $\Lambda_{(27)}^4$ component of $\tr(R^\kappa\wedge R^\kappa)$ cannot vanish for any choice of $\kappa$. This fact will pose severe constraints on the possible ans\"atze for the gauge field.

\section{The heterotic string on \np \gmfs}
\label{sec:hetStr_on_adsxX7}

In this section we discuss the solutions of the field equations and Bianchi identity of the heterotic string on
\begin{equation}
  M = M_{1,2} \times X_7 \:,
\end{equation}
with $M_{1,2}$ being three-dimensional de Sitter, anti-de Sitter or Minkowski space, and $X_7$ being a \np \gmf or a manifold with \ghol.  Moreover, we calculate the supersymmetry variations and the masses of the fermions for any of these solutions. 

As the gaugino and dilatino condensates possess components on both $M_{1,2}$ and $X_7$, the vanishing of the dilatino supersymmetry variation \eqref{eq:dilatinoVariation} demands that the same holds true for the three-form flux $H$.  Hence, it is natural to choose 
\begin{equation}
  H = -h_1 \; \vol3 +\; h_2 \; Q 
\end{equation}
with $h_1,\,h_2 \in \RR$, as an ansatz also for not necessarily supersymmetric solutions.

The possible ans\"atze for the gauge field $F$ and the connection $\tilde \Gamma$ are highly restricted by the Bianchi identity.  Here, we choose $F$ to be the curvature of the canonical connection $\nabla^C$ on $X_7$.  Although it is in principle possible to find non-instanton solutions, instanton solutions are distinguished by allowing for a non-standard embedding in the Bianchi identity and by immediately solving the Yang-Mills equation.  Furthermore, for an instantonic gauge field, supersymmetry variation of the gaugino vanishes.  We identify the curvature $\tilde R$ of the connection $\tilde \Gamma$ with the curvature of the interpolating connection $\nabla^\kappa$.  As discussed in Section \ref{sec:g2manifolds}, this choice ensures the vanishing of the $\Lambda_{(27)}^4$ component of the right hand side of the Bianchi identity \eqref{eq:BianchiEQ}, which is required as its left hand side, $\dd H$, is proportional to $*Q \in \Lambda_{(1)}^4$. Summarizing, our ansatz is
\begin{equation}
F=R^C \und \tilde R=R^\kappa \;.
\end{equation}

\subsection{Equations of motion} 
\label{sec:equations-motion}
For our ansatz, the equations of motion \eqref{eq:10D-eom} reduce to a set of algebraic equations for the parameters $\mu_1, \mu_2, h_1, h_2, m, n$ and $\kappa$.  It is convenient to replace the parameters $m$ and $n$ by $\hat m$ and $\hat n$, defined as
\begin{equation}
  2\, \hat m = 4n - m \und
  2\, \hat n = 3m - 4n \:.
\end{equation}
The Einstein equation splits into one equation on $M_{1,2}$ and one on $X_7$:
\begin{subequations}
\label{eq:EOM_AnsatzInserted}
\begin{align}
\label{eq:EinsteinEQExt_AnsatzInserted}
 64\mu_1^2(1+2\ap\mu_1^2) - 2 h_1(h_1+\hat m) &= 0 \:, \\
\label{eq:EinsteinEQInt_AnsatzInserted}
 64\,\mu_2^2(1- \tfrac{2}{27} \ap \mu_2^2 a(\kappa)) - 2 h_2(h_2+ \hat m)  &= 0 \:,
\end{align}
with
\begin{equation}
 a(\kappa) = \kappa^2(16-11\kappa^2) \:.
\end{equation}
Furthermore, the dilaton equation and the Bianchi identity read
\begin{equation}
\label{eq:DilatonEQ_AnsatzInserted}
 -48\mu_1^2(1-\ap\mu_1^2) + 336\mu_2^2 - \tfrac{112}9 \ap \mu_2^4\, a(\kappa) - h_1^2 - 7 h_2^2 - 2 (h_1 +7 h_2) \hat{m} + \hat{m}^2 -\hat{n}^2 = 0
 \end{equation}
\end{subequations}
and
\begin{equation}
 \label{eq:BianchiIdentity_AnsatzInserted}
 \mu_2 h_2 = \frac{8}{27} \mu_2^4 \,\ap\, b(\kappa)
\end{equation}
with
\begin{equation}
 b(\kappa) = \kappa^2 (1+\kappa^2) \:,
\end{equation}
respectively.  Note that $\hat n$ enters the equations of motions only quadratically.  Hence, solely $\hat n^2$ will be determined by the field equations, but the sign of $\hat n$ will not be fixed.

\subsection{Solution to the equations of motion}

\subsubsection{De Sitter and anti-de Sitter solutions}

\paragraph{\Np $G_2$ compactifications.} 
We begin with the more general situation of compactifications on \np \gmfs, i.e.~we assume $\mu_2\neq0$.  As discussed in Section \ref{sec:g2manifolds}, we set $\mu_2=\rho_2^{-1}$ in this case, where $|\rho_2|$ measures the size of $X_7$.  As the solutions for de Sitter and anti-de Sitter backgrounds are very similar, we treat them in parallel.  We remind the reader that the radius of the (anti-) de Sitter space is given by $|\rho_1|=|\mu_1|^{-1}$, with $\mu_2$ being either real or purely imaginary.
The equations of motion \eqref{eq:EOM_AnsatzInserted} and the Bianchi identity \eqref{eq:BianchiIdentity_AnsatzInserted} form a system of four algebraic equations for the seven parameters of our ansatz, 
\begin{equation}
\rho_1, \rho_2, \kappa, h_1, h_2, \hat m \ \textrm{and}\ \hat n \:.
\end{equation}
It can be solved for the parameters of the $H$-flux and the fermionic condensates,
\begin{equation}
h_1(\rho_1,\rho_2,\kappa)\:,\quad h_2(\rho_1,\rho_2,\kappa)\:,\quad
\hat{m}(\rho_1,\rho_2,\kappa)\:,\quad \hat{n}(\rho_1,\rho_2,\kappa)\:.
\end{equation}
The explicit expressions are as follows,
\begin{subequations}
\begin{align}
  \label{eq:EOMsol_h1_ads}
  \begin{split}
  h_1^\pm &= -\rho_2\left[\frac{54}{\ap b(\kappa)}-\frac{4 a(\kappa)}{\rho_2^2 b(\kappa)}-\frac{4\ap  b(\kappa)}{27 \rho_2^4}\right]
             \pm \rho_2 \sqrt{\left[\frac{54}{\ap b(\kappa)}-\frac{4 a(\kappa)}{\rho_2^2 b(\kappa)}-\frac{4\ap  b(\kappa)}{27 \rho_2^4}\right]^2 + \frac{32\left(2\ap +\delta\rho_1^2\right)}{ \rho_1^4\rho_2^2}} \:,
  \end{split}\\
  \label{eq:EOMsol_m_ads}
  h_2 &= \frac{8 \ap}{27 \rho_2^3} b(\kappa) 
\end{align}
and
\begin{align}
  \hat m &= 2\rho_2 \left[\frac{54}{\ap b(\kappa)}-\frac{4 a(\kappa)}{\rho_2^2 b(\kappa)}-\frac{4\ap  b(\kappa)}{27 \rho_2^4}\right] \:, \\
  \label{eq:EOMsol_n_ads}
  \begin{split}
    (\hat n^\pm)^2 &= 
      -16\frac{\ap-\delta\rho_1^2}{\rho_1^4}
      +16\left[
        \frac{2187 \rho_2^2}{2 \ap^2 b(\kappa)^2}
        -\frac{162 a(\kappa)}{\ap b(\kappa)^2}
        -\frac{13}{\rho_2^2}
        +\frac{6 a(\kappa)^2}{\rho_2^2 b(\kappa)^2}
        +\frac{47\ap a(\kappa)}{27\rho_2^4}
        +\frac{34 \ap^2 b(\kappa)^2}{729 \rho_2^6}
      \right] \\
      &\quad
      \pm2\rho_2^2\left[\frac{54}{\ap b(\kappa)}-\frac{4 a(\kappa)}{\rho_2^2 b(\kappa)}-\frac{4\ap b(\kappa)}{27 \rho_2^4}\right]
        \sqrt{
          \left[\frac{ 54}{\ap b(\kappa)}-\frac{4 a(\kappa)}{\rho_2^2 b(\kappa)}-\frac{4\ap b(\kappa)}{27 \rho_2^4}\right]^2
          +\frac{32(2 \ap+\delta\rho_1^2) }{\rho_1^4 \rho_2^2}
        } \:,
   \end{split}
\end{align}
\end{subequations}
respectively, with the plus/minus signs in \eqref{eq:EOMsol_h1_ads} and \eqref{eq:EOMsol_n_ads} being correlated.  We have also defined
\begin{equation}
  \delta =
  \begin{cases}
    -1 & \text{for anti-de Sitter backgrounds} \\
    1 & \text{for de Sitter backgrounds}
  \end{cases}\:.
\end{equation}
For fixed values of $\rho_1$, $\rho_2$ and $\kappa$, the field equations possess up to four solutions for the parameters $h_1$, $h_2$, $\hat m$ and $\hat n$, distinguished by the choice of the plus/minus-signs in \eqref{eq:EOMsol_h1_ads} and \eqref{eq:EOMsol_n_ads} and the sign of $\hat n$.  However, the equations of motion are not solvable for all values of $(\rho_1, \rho_2, \kappa)$.  The excluded regions in the parameter space $(\rho_1, \rho_2, \kappa)$ are discussed in Appendix \ref{sec:restr-param-space}.

\begin{figure}[h]
  \centering
  \begin{subfigure}[b]{.45\textwidth}
    \hspace{.4cm}
    \includegraphics[width=\textwidth]{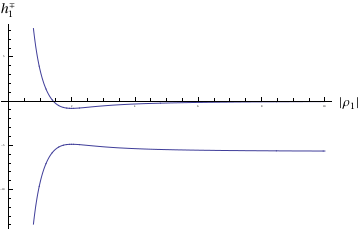}
    \caption{Anti-de Sitter backgrounds}
  \end{subfigure}
  \hfill
  \begin{subfigure}[b]{.45\textwidth}
    \hspace{.4cm}
    \includegraphics[width=\textwidth]{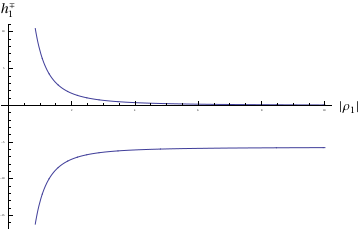}
    \caption{De Sitter backgrounds}
  \end{subfigure}
  \caption{Plots of $h_1^\pm$ for a fixed value of $\rho_2$ and $\kappa$.}
  \label{fig:plot_h1_rho1}
\end{figure}

\begin{figure}[h]
  \centering
  \includegraphics[width=.45\textwidth]{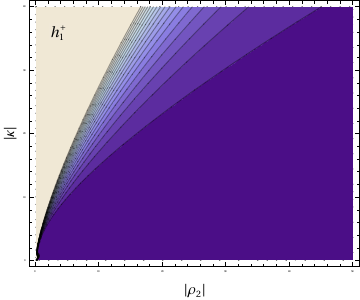}
  \includegraphics[width=.45\textwidth]{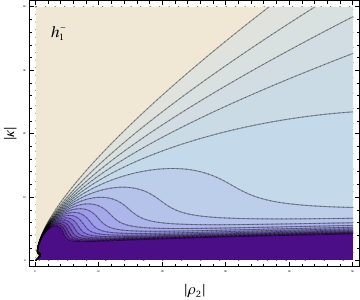}
  \caption{Contour plots of $h_1^\pm$ for a fixed value of $\rho_1$.}
  \label{fig:plot_h1_rho2kappa}
\end{figure}

\begin{figure}[h]
  \centering
    \includegraphics[width=.45\textwidth]{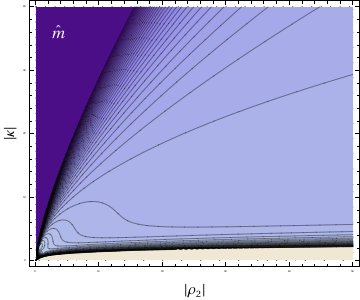}
    \caption{Contour plot of $\hat m$.}
    \label{fig:plot m}
\end{figure}

\begin{figure}[h]
  \centering
  \begin{subfigure}[b]{.45\textwidth}
    \hspace{.4cm}
    \includegraphics[width=\textwidth]{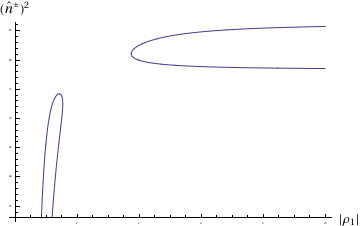}\\[.5cm]
    \includegraphics[width=\textwidth]{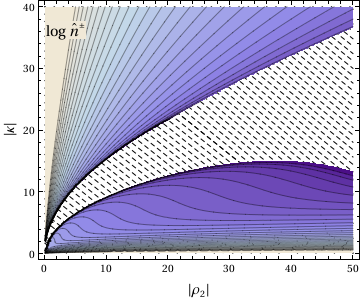}
    \caption{Anti-de Sitter backgrounds}
  \end{subfigure}
  \hfill
  \begin{subfigure}[b]{.45\textwidth}
    \hspace{.4cm}
    \includegraphics[width=\textwidth]{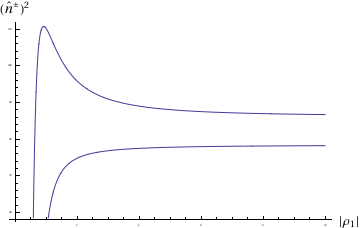}\\[.5cm]
    \includegraphics[width=\textwidth]{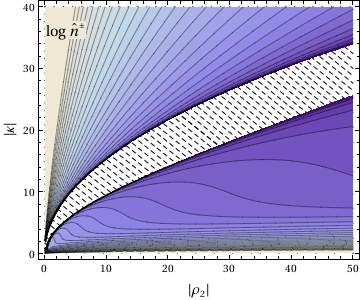}
    \caption{De Sitter backgrounds}
  \end{subfigure}
  \caption{Plots of $\hat n^\pm$ for a fixed value of $\rho_2$ and $\kappa$ (top) and contour plots of $\hat n^-$ (bottom) for a fixed value of $\rho_1$.  In the striped area, no solution to the equations of motion exist.}
  \label{fig:plot n}
\end{figure}

The dependence of $h_1$, $\hat{m}$ and $\hat{n}$ on the parameters $\rho_1$, $\rho_2$ and $\kappa$ cannot be read off easily from \eqref{eq:EOMsol_h1_ads}, \eqref{eq:EOMsol_m_ads} and \eqref{eq:EOMsol_n_ads}.  It is depicted qualitatively in the Figures \ref{fig:plot_h1_rho1}, \ref{fig:plot_h1_rho2kappa}, \ref{fig:plot m} and \ref{fig:plot n}.

\paragraph{Solutions with $h_1=0$ or $h_2=0$.} 
In general, the tensor field $H$ has components proportional to $Q$ as well as to $\vol3$.  We remark that there is no solution with $h_2=0$: In this case, the Bianchi identity \eqref{eq:BianchiIdentity_AnsatzInserted} implies $\kappa=0$, but obviously the Einstein equation \eqref{eq:EinsteinEQInt_AnsatzInserted} possesses no solution with $h_2=\kappa=0$.  On the other hand, $h_1=0$ is possible: From \eqref{eq:EinsteinEQExt_AnsatzInserted} we can read off that, in anti-de Sitter backgrounds, $h_1=0$ only implies $\rho_1^2=2\ap$.  Since $h_2$ and $\hat m$ do not depend on $\rho_1$, their expressions are not simplified for this special solution.  However, \eqref{eq:EOMsol_n_ads} reduces to 
\begin{equation}
  \label{eq:h1=0 EOMsol_n}
  (\hat n)^2 = 
    -\frac{12}{\ap}
    -\frac{112}{\rho_2^2}
    +\frac{560 \ap k(\kappa)}{27\rho_2^4}
    +\frac{448 \ap^2 b(\kappa)^2}{729\rho_2^6}
    +4 \rho_2^2 \left[ \frac{54}{\ap b(\kappa)} - \frac{4 k(\kappa)}{\rho_2^2 b(\kappa)} - \frac{4\ap b(\kappa)}{27 \rho_2^4} \right]^2 \:.
\end{equation}

\begin{figure}[h]
  \centering
  \begin{subfigure}[b]{.45\textwidth}
    \includegraphics[width=\textwidth]{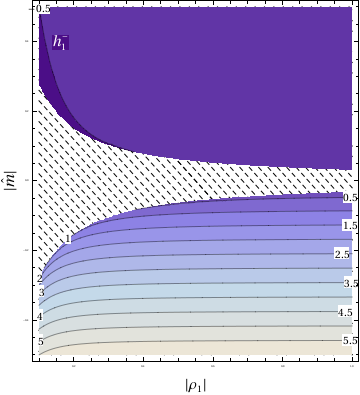}
    \caption{Anti-de Sitter background}
  \end{subfigure}
  \hfill
  \begin{subfigure}[b]{.45\textwidth}
    \includegraphics[width=\textwidth]{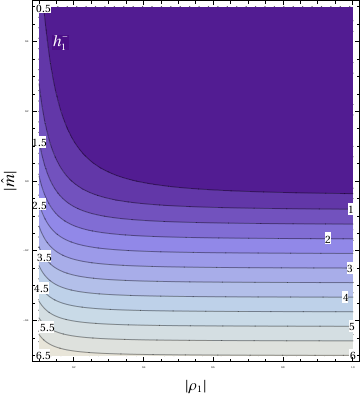}
    \caption{De Sitter background}
  \end{subfigure}
  \caption{Contour plots of $h_1^-$ for compactifications on manifolds with \ghol.}
  \label{fig:plot_h1_g2hol}
\end{figure}

\begin{figure}[h]
  \centering
  \begin{subfigure}[b]{.45\textwidth}
    \includegraphics[width=\textwidth]{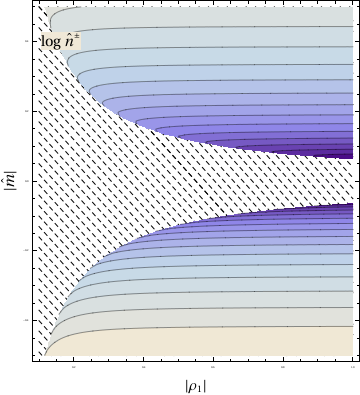}
    \caption{Anti-de Sitter background}
  \end{subfigure}
  \hfill
  \begin{subfigure}[b]{.45\textwidth}
    \includegraphics[width=\textwidth]{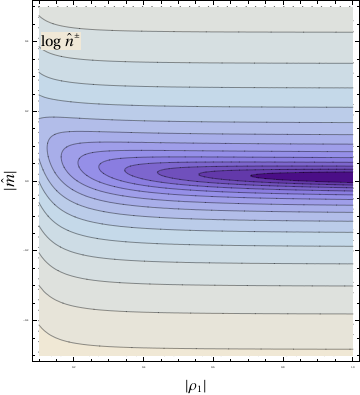}
    \caption{De Sitter background}
  \end{subfigure}
  \caption{Plots of $\hat n^\pm$ for compactifications on manifolds with \ghol.  In the striped area, no solution to the equations of motion exist.}
  \label{fig:plot_n_g2hol}
\end{figure}

\paragraph{\ghol compactifications.}
If we choose, more specially, $X_7$ to have \ghol, the Bianchi identity can only be solved by the standard embedding,
\begin{equation}
F=\tilde R \qquad\Rightarrow\qquad \kappa=0\:.  
\end{equation}
The reason is that, on \ghol manifolds, the three form $Q$, and thus $H$, is closed, and thereby the left-hand side of the Bianchi identity \eqref{eq:BianchiEQ} vanishes. Furthermore, since $\mu_2=0$, we send $\rho_2\to\infty$. This leaves us with 3 equations for 5 parameters, which we can solve for
\begin{equation}
h_1(\rho_1,\hat m)\ ,\quad h_2(\rho_1,\hat m) \und \hat n(\rho_1,\hat m)\:.
\end{equation}
The Einstein equation on \adsds3 \eqref{eq:EinsteinEQExt_AnsatzInserted} is solved by
\begin{subequations}
\label{eq:EOMsol_g2hol}
\begin{equation}
  \label{eq:EOMsol_h1_ads_g2hol}
  h_1^\pm = -\frac{\hat m}2 \pm\frac1{\rho_1^2}\sqrt{32(2\ap+\delta\rho_1^2)+\tfrac14\hat m\rho_1^4} \:.
\end{equation}
The Einstein equation on $X_7$ \eqref{eq:EinsteinEQInt_AnsatzInserted} admits only two solutions,
which we distinguish by introducing an auxiliary parameter~$\theta$:
\begin{alignat}{2}
h_2 &= 0 & \qquad\Leftrightarrow\qquad \theta &= 3 \oder \\[-.1cm]
h_2 &= -\hat m & \qquad\Leftrightarrow\qquad \theta &= 17 \:.
\end{alignat}
Finally, the dilaton equation \eqref{eq:DilatonEQ_AnsatzInserted} is solved by 
\begin{equation}
  \label{eq:EOMsol_n_ads_g2hol}
  (\hat n^\pm)^2 = 16\frac{\delta\rho_1^2+\ap}{\rho_1^4} + \frac\theta2\hat m^2 \pm\frac{\hat m}{\rho_1^2}\sqrt{32(2\ap+\delta\rho_1^2)+\tfrac14\hat m^2\rho_1^4} \:,
\end{equation}
\end{subequations}
with the plus/minus signs in \eqref{eq:EOMsol_h1_ads_g2hol} and \eqref{eq:EOMsol_n_ads_g2hol} being correlated.
Thus, in contrast to \np $G_2$ compactifications, there are up to eight solutions to the field equations, parametrized by $\rho_1$ and $\hat m$ and distinguished by the choice of the plus/minus-sign in \eqref{eq:EOMsol_h1_ads_g2hol} and \eqref{eq:EOMsol_n_ads_g2hol}, the parameter $\theta$ and the sign of $\hat n$.  The qualitative dependence of $h_1$ and $\hat n$ on $\rho_1$ and $\hat m$ are depicted in Figures \ref{fig:plot_h1_g2hol} and \ref{fig:plot_n_g2hol}.  Unlike in the case of \np $G_2$ compactifications, there now exist solutions with $H$-flux confined to $M_{1,2}$, i.e.~$h_2=0$ and $h_1\neq 0$.  

As for \np $G_2$ compactifications, the solutions on anti-de Sitter backgrounds simplify for \mbox{$\rho_1^2=2\ap$}:
\begin{align}
  h_1^\pm &=  \frac{-\hat m \pm|\hat m|}{2} \:, \\
  (\hat n^\pm)^2 &= \frac{12}{\ap} +\frac{\hat m(\theta\,\hat m \pm |\hat m|)}{2} \:.
\end{align}
Obviously, this yields also solutions with vanishing flux on $M_{1,2}$,
\begin{equation}
  h_1 = 0 \und
  \hat n^2 = \frac{12}{\ap} + \frac{(1+\theta)\hat m^2}2 \:.
\end{equation}
Moreover, there is also a solution with completely vanishing $H$-flux and condensates only,
\begin{equation}
  h_1 = h_2 = 0 \und \hat n^2 = \frac{12}{\ap} + 2\hat m^2 \:.
\end{equation}

\subsubsection{Minkowski solutions}

\paragraph{\Np $G_2$ compactifications.}
For compactifications to three-dimensional Minkowski space, we have to set $\mu_1=0$.  The solutions to the Bianchi identity \eqref{eq:BianchiIdentity_AnsatzInserted} and the Einstein equation on $X_7$ \eqref{eq:EinsteinEQInt_AnsatzInserted} are the same as in the (A)dS-case,
\begin{subequations}
\begin{align}
  \label{eq:EOMsol_m_mink}
  h_2 &= \frac{8 \ap}{27 \rho_2^3} b(\kappa) \:.\\
  \hat m &= 2\rho_2 \left[\frac{54}{\ap b(\kappa)}-\frac{4 a(\kappa)}{\rho_2^2 b(\kappa)}-\frac{4\ap  b(\kappa)}{27 \rho_2^4}\right] \:.
\end{align}
The Einstein equation on $M_{1,2}$ \eqref{eq:EinsteinEQExt_AnsatzInserted} on the other hand is solved by either
\begin{align}
 \label{eq:EOMsol_h1_mink+}
  h_1^- &= 0 \oder \\ 
 \label{eq:EOMsol_h1_mink-}
  h_1^+ &= -2\rho_2 \left[\frac{54}{\ap b(\kappa)}-\frac{4 a(\kappa)}{\rho_2^2 b(\kappa)}-\frac{4\ap  b(\kappa)}{27 \rho_2^4}\right] \:.
\end{align}
Finally, the solution to the equation of motion for the dilaton \eqref{eq:DilatonEQ_AnsatzInserted} yields
\begin{equation}
  \label{eq:EOMsol_n_mink}
  (\hat n^\pm)^2 = 
    112\left(
      -\frac1{\rho_2^2}
       +\frac{5\ap a(\kappa)}{27\rho_2^4}
       +\frac{4\ap^2 b(\kappa)^2}{729\rho_2^6}
     \right)
     +(6 \pm 2) \rho_2^2 \left(
       \frac{54}{\ap b(\kappa)}
       -\frac{4 a(\kappa)}{\rho_2^2 b(\kappa)}
       -\frac{4\ap b(\kappa)}{27 \rho_2^4}
     \right)^2 \:.
\end{equation}
\end{subequations}
The superscripts $+$ and $-$ refer to the solutions \eqref{eq:EOMsol_h1_mink+} and \eqref{eq:EOMsol_h1_mink-} for $H_1$.  As for the solutions on de Sitter and anti-de Sitter backgrounds, the dependence of $\hat m$ and $\hat n$ on $\rho_2$ and $\kappa$ cannot be easily seen from \eqref{eq:EOMsol_m_mink} and \eqref{eq:EOMsol_n_mink}.  The solution for $\hat m$, however, is identical to the (anti-)de Sitter case (see \eqref{eq:EOMsol_m_ads} and Figure \ref{fig:plot m}).  Furthermore, the qualitative dependence of the solution for $\hat n$ on $\rho_2$ and $\kappa$ is the same as for (anti-)de Sitter backgrounds (see Figure \ref{fig:plot n}), as \eqref{eq:EOMsol_n_mink} is the $|\rho_1|\rightarrow\infty$ limit of \eqref{eq:EOMsol_n_ads}.

\paragraph{\ghol compactifications.}
Finally, we discuss the case of compactifications to Minkowski space on manifolds with \ghol.  As in the previous cases of \ghol compactifications, the Bianchi identity requires
\begin{align}
  \tilde R = F \qquad\Rightarrow\qquad \kappa=0\:.
\end{align}
After sending $\rho_2\to\infty$, the field equations are solved by
\begin{subequations}
\begin{equation}
  h_2 = -\hat m
\end{equation}
and either
\begin{alignat}{2}
  h_1 &= 0 \:,\quad &
  \hat n^2 &= 8 \hat m^2
\oder\\
  h_1 &= -\hat m \:,\quad &
  \hat n^2 &= 9 \hat m^2 \:.
\end{alignat}
\end{subequations}
Hence, there are four independent solutions to the field equations.  In contrast to the previous cases, the equations of motion possess solutions for all values of the parameter $\hat m$.

\subsection{Supersymmetry conditions and supersymmetric solutions}
\label{sec:susy-cond}
It is of interest to find out which subset of our heterotic backgrounds are supersymmetric. To this end, we investigate the supersymmetry conditions \eqref{eq:SUSYconditions}. We begin with the gravitino variation \eqref{eq:gravitinoVariation}.  Recall that $M_{1,2}$ carries either a real or imaginary Killing spinor $\zeta$ with Killing number $\mu_1$ and that $X_7$ possesses a real Killing spinor with Killing number $\mu_2$.  Furthermore, using \eqref{eq:Psi.eta} and \eqref{eq:X_Psi.eta} it is straightforward to compute
\begin{equation}
 (X \lrcorner H) \cdot \epsilon =
 \begin{cases}
  -2 \, h_1 \: ( X  \cdot \widehat\epsilon ) \otimes \eta \otimes (1,0)^t & \text{for }  X\in TM_{1,2} \\[.2cm]
  \quad 6 i \, h_2 \: \epsilon \otimes ( X \cdot \eta ) \otimes (1,0)^t & \text{for } X\in TX_7
 \end{cases}
\end{equation}
and
\begin{equation}
 \Sigma \cdot X \cdot \epsilon =
  \begin{cases}
   -m \;\; (X \cdot \widehat\epsilon) \otimes \eta \otimes (1,0)^t & \text{for }  X\in TM_{1,2} \\[.2cm]
   - 7 i\, m \;\; \widehat\epsilon \otimes (X \cdot \eta) \otimes (1,0)^t & \text{for } X\in TX_7
  \end{cases} \;.
\end{equation}
Inserting these relations in the gravitino variation yields
\begin{subequations}
\label{eq:gravitino variation,ansatz inserted}
\begin{alignat}{2}
  96 \mu_1 &= 12 h_1 + m    &&= 12 h_1 + \hat m + \hat n\:, \\
  96 \mu_2 &= 12 h_2 + 7 m  &&= 12 h_2 + 7(\hat m + \hat n) \:.
\end{alignat}
\end{subequations}
As we set the dilaton to zero, the dilatino variation \eqref{eq:dilatinoVariation} reads
\begin{align}
  4 H + \Sigma - \Delta = 0
\end{align}
and therefore we obtain
\begin{equation}
  \label{eq:dilatino variation, ansatz inserted}
  16 h_1 = 16 h_2 = 4(n-m) = -\hat m - 3\hat n\:.
\end{equation}
As already mentioned at the beginning of this section, the gaugino variation \eqref{eq:gauginoVariation} vanishes since $F$ is an instanton.

Together with the equations of motion, the conditions for the vanishing of the gravitino and the dilatino variation, \eqref{eq:gravitino variation,ansatz inserted} and \eqref{eq:dilatino variation, ansatz inserted}, form a set of eight equations for the seven parameters $\mu_1$, $\mu_2$, $h_1$, $h_2$, $\hat m$, $\hat n$ and $\kappa$.  It can be checked straightforwardly that this system is not solvable.  Hence, our compactifications on \np \gmfs do not yield any supersymmetric solutions for either de Sitter, anti-de Sitter or Minkowski space-times, not even in the \ghol limit.

\subsection{Fermion masses}
\label{sec:fermion-masses}
Employing the decomposition of the fermions \eqref{eq:decomposition_of_fermions} and the Killing property of $\eta$ as well as the knowledge of the Clifford action of $\vol3$ \eqref{eq:clifford_action_of_vol3} and $Q$ \eqref{eq:Psi.eta}, it is straightforward to rewrite the Dirac equations for the gaugino \eqref{eq:dirac eq gaugino} and dilatino \eqref{eq:dirac eq dilatino} as
\begin{align}
 0 &= \left( \tilde{\mathcal D} + \mathcal D^- - \tfrac 1{24} (H-\tfrac 12\Sigma+\tfrac 12\Delta)\cdot \right) \: \chi = (\tilde{\mathcal D} - M_\chi ) \: \chi \:, \\
 0 &= \left( \tilde{\mathcal D} + \mathcal D^- - \tfrac 1{24} (H-\tfrac 18\Sigma) \cdot \right) \: \lambda = (\tilde{\mathcal D} - M_\lambda ) \: \lambda \:,
\end{align}
where $\tilde{\mathcal D}$ is the Dirac operator on $M_{1,2}$.  The masses of the gaugino and dilatino are readily computed to be
\begin{equation}
  M = \frac{1}{24} \left(h_1+7 h_2-\hat{m}-c\hat{n}\right) 
  \qquad\textrm{with}\qquad c_\chi = 3 \und c_\lambda = 1\:.
\end{equation}
In the following, we give the expressions of the masses in terms of the parameters $(\rho_1,\rho_2,\kappa)$ or $(\rho_1,\hat m)$, respectively, for the compactifications considered previously.

\paragraph{De Sitter and anti-de Sitter backgrounds.}
Recall that for given values of $(\rho_1,\rho_2,\kappa)$ the field equations possess up to four solutions in the case of compactifications on \np \gmfs and up to eight solutions for compactifications on manifolds with \ghol.  There are up to two solutions for $h_1$ labeled by the superscript $\pm$.  Then, for a given solution for $h_1$, the field equations fix the value of $\hat n^2$, but not the sign of $\hat n$ (see \eqref{eq:EOMsol_n_ads}).  Hence, in the following we set
\begin{equation}
  \hat n = \nu\, |\hat n| \qquad\textrm{with}\qquad \nu \in \{ -1 , +1 \}\:.
\end{equation}

For compactifications on \np \gmfs the fermion masses are given by
\begin{align}
  M^\pm
   &=-\frac{27 \rho_2}{4 \ap b(\kappa) }
     +\frac{a(\kappa)}{2 \rho_2 b(\kappa)}
     +\frac{17 \ap b(\kappa)}{162 \rho_2^3 }
     \mp\frac{\rho_2}{12} 
        \sqrt{
           \frac{8\left(2 \ap +\delta\rho_1^2\right)}{\rho_1^4 \rho_2^2}
          +\left(\frac{27}{\ap b(\kappa)} - \frac{2 a(\kappa)}{\rho_2^2 b(\kappa)} - \frac{2\ap b(\kappa)}{27\rho_2^4} \right)^2 
        }\notag\\
     &\quad
      -\frac{\nu c|\rho_2|}{6\sqrt2} 
         \left[
           \pm\left(\frac{27}{\ap b(\kappa)} - \frac{2 a(\kappa)}{\rho_2^2 b(\kappa)} - \frac{2\ap b(\kappa)}{27\rho_2^4} \right)
             \sqrt{
               \frac{8\left(2 \ap +\delta\rho_1^2\right)}{\rho_1^4 \rho_2^2}
               +\left( \frac{27}{\ap b(\kappa)} - \frac{2 a(\kappa)}{\rho_2^2 b(\kappa)} - \frac{2\ap b(\kappa) }{27\rho_2^4} \right)^2
             }
            \right.\notag\\
            &\quad\qquad\qquad\qquad\left.
            \frac{2 \left(\delta\rho_1^2 + \ap\right)}{\rho_1^4\rho_2^2}
           +\frac{3 \left(27 \rho_2^2 - 2 \ap a(\kappa) \right)^2}{\ap^2 b(\kappa)^2 \rho_2^4} 
           -\frac{26}{\rho_2^4}
           +\frac{94 \ap a(\kappa)}{27\rho_2^6}
           +\frac{68 \ap^2 b(\kappa)^2}{729\rho_2^8}
        \right]^{1/2}\:,
\end{align}
whereas for compactifications on manifolds with \ghol they read\footnote{Recall, that $\theta=3$ for solutions with $h_2=0$ and $\theta=17$ for $h_2=-\hat m$ (see \eqref{eq:EOMsol_g2hol}).}
\begin{align}
  \notag
  M^\pm &= 
   -\theta\frac{\hat{m}}{48}
   \pm\frac1{48 \rho_1^4}\sqrt{128\rho_1^4 (\delta \rho_1^2+2 \ap )+\rho_1^8 \hat{m}^2}\\
   &\quad-\frac{c\nu}{24 \sqrt{2} \rho_1^2} \sqrt{ 32 \left(\delta \rho_1^2-\ap \right)+\theta \rho_1^4 \hat{m}^2 \pm \rho_1^2 \hat{m} \sqrt{128 (\delta \rho_1^2 +2 \ap )+\rho_1^4 \hat{m}^2}} \:.
\end{align}

\paragraph{Minkowski backgrounds.}
On Minkowski backgrounds, the masses of the gaugino and dilatino are given by
\begin{align}
  \notag
  M^\pm &= 
    \frac{(6\pm2) k(\kappa)}{3\rho_2 b(\kappa)}
   -\frac{9(6\pm2)\rho_2}{2\ap b(\kappa)}
   +\frac{(13\pm2)\ap b(\kappa)}{81 \rho_2^3}\\
   &\quad
   -\frac{c\nu}{8} \sqrt{
    112\left(
      -\frac1{\rho_2^2}
       +\frac{5\ap a(\kappa)}{27\rho_2^4}
       +\frac{4\ap^2 b(\kappa)^2}{729\rho_2^6}
     \right)
     +(6\pm2) \rho_2^2 \left(
       \frac{54}{\ap b(\kappa)}
       -\frac{4 a(\kappa)}{\rho_2^2 b(\kappa)}
       -\frac{4\ap b(\kappa)}{27 \rho_2^4}
     \right)^2
    }
\end{align}
for compactifications on \np \gmfs.

Finally, for compactifications on manifolds with \ghol the fermion masses read
\begin{align}
  M^\pm &= -\tfrac1{24}(a\, \hat m + c\, b\, \nu |\hat m| ) \:.
\end{align}
The values of $(a,b)$ are given in the following table for all solutions to the field equations:
\begin{align}
  \begin{array}{l||l|l}
    (a,b) & h_1 = 0 & h_1 = -\hat m \\[.05cm]
    \hline
 \hline
    h_2 = 0 & (1,\,1) & (2,\,\sqrt2) \\
    \hline
    h_2 = \hat m & (4,\,2\sqrt2) & (9,\,3) 
  \end{array}
\end{align}

\section*{Acknowledgements}
We thank Alexander Haupt and Alexander Popov for helpful comments.  This work was supported in part by Deutsche Forschungsgemeinschaft grant LE 838/13 and the Graduiertenkolleg 1463 ``Analysis, Geometry and String Theory``.  Furthermore, Karl-Philip Gemmer thanks the Volkswagen Foundation for partial support under the grant I/84 496.

\clearpage

\begin{appendix}

\section{Useful identities for \np \gmfs}
\label{sec:app:npg2}
In this appendix we will list some useful identities for the $G_2$-invariant three-form $Q$ on manifolds with \gstr.  In the following we will denote the Hodge-dual of $Q$ by $\hq$,
\begin{equation}
  \hq \equiv * Q = e^{3456} + e^{2467} + e^{2357} + e^{1457} + e^{1367} + e^{1256} + e^{1234} \:.
\end{equation}
It is straightforward to derive the following identities of contractions of $Q$ and $\hq$ \cite{Bilal2001}:
\begin{subequations}
\label{eq:psi-identities}
\begin{align}
Q_{abe}Q_{cde} &= -\hq_{abcd}+\delta_{ac}\delta_{bd} -\delta_{ad}\delta_{bc} \,, \\
Q_{acd}Q_{bcd} &= 6\, \delta_{ab} \,, \\
Q_{abp}\hq_{pcde} &= 3 Q_{a[cd}\delta_{e]b} - 3Q_{b[cd}\delta_{e]a} \,, \\
\hq_{abcp}\hq_{defp} &= -3\hq_{ab[de}\delta_{f]c}-2\hq_{def[a}\delta_{b]c} -3Q_{ab[d}Q_{ef]c}
+ 6\delta^{[d}_a\delta^e_b\delta^{f]}_c \,, \\
\hq_{abpq}Q_{pqc} &= -4Q_{abc} \,, \\
\hq_{abpq}\hq_{pqcd} &= -2\hq_{abcd} + 4(\delta_{ac}\delta_{bd} - \delta_{ad}\delta_{bc}) \,, \\
\hq_{apqr}\hq_{bpqr} &= 24\delta_{ab} \,, \\
Q_{abp}Q_{pcq}Q_{qde} &= Q_{abd}\delta_{ce} - Q_{abe}\delta_{cd} - Q_{ade}\delta_{bc} + Q_{bde}\delta_{ac} \nonumber\\
&\qquad - Q_{acd}\delta^{be} + Q_{ace}\delta_{bd} + Q_{bcd}\delta_{ae} - Q_{bce}\delta_{ad} \,, \\
\label{ident10}
 Q_{paq}Q_{qbs}Q_{scp} &= 3Q_{abc} \, .
\end{align}
\end{subequations}

\section{Restrictions on the parameter space of solutions}
\label{sec:restr-param-space}
In general, the equations of motion do not possess solutions for all values of the parameters $\mu_1, \mu_2, h_1, h_2, \hat m, \hat n$ and $\kappa$ of the ansatz considered in Section \ref{sec:hetStr_on_adsxX7}.  In this appendix we discuss for which values of the parameters the field equations are not solvable.

\subsection{\Np $G_2$ compactifications}
\begin{figure}[!h]
  \begin{center}
    \begin{subfigure}[b]{.45\textwidth}
      \includegraphics[width=\textwidth]{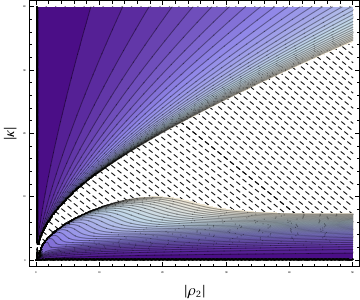}
      \caption{Anti-de Sitter backgrounds}
    \end{subfigure} 
    \begin{subfigure}[b]{.45\textwidth}
      \includegraphics[width=\textwidth]{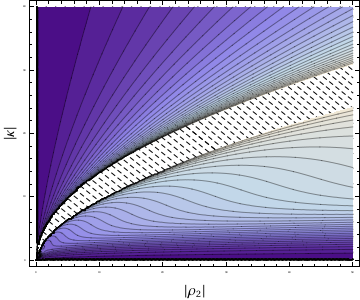}
      \caption{De Sitter backgrounds}
    \end{subfigure}
  \end{center}
  \caption{Contour plot of the lower bound for $|\rho_1|$ for which solutions to the equations of motion for (anti-)de Sitter backgrounds exists.  For $\rho_2 \rightarrow 0$ or $\kappa \rightarrow 0$ the lower bound approaches zero.  In the striped area, the equations of motion cannot be solved for any value of $\rho_1$.}
  \label{fig:restriction on parameter space_npg2_adsds}
\end{figure}

For both, anti-de Sitter and de Sitter backgrounds, the equations of motion were solved for $h_1, h_2, \hat m$ and $\hat n$ and are parametrized by $\rho_1 = \mu_1^{-1}$, $\rho_2=\mu_2^{-1}$ and $\kappa$.  In both cases it is obvious that there exists a lower bound for $|\rho_1|$ imposed by demanding the solution for $\hat n^2$, \eqref{eq:EOMsol_n_ads} or \eqref{eq:EOMsol_n_ads}, respectively, to be positive.  The same conditions exclude for fixed values of $\rho_1$ a region in the parameter space spanned by $\rho_2$ and $\kappa$ in which no solution to the field equations exists.  Additionally, in the case of anti-de Sitter backgrounds, the argument in the square root in \eqref{eq:EOMsol_h1_ads} and \eqref{eq:EOMsol_n_ads} becomes negative for certain values of $\rho_1$, $\rho_2$ and $\kappa$.  Contour plots of the resulting lower bound on $|\rho_1|$ and the excluded region in the parameter space spanned by $\rho_2$ and $\kappa$ are depicted in Figure \ref{fig:restriction on parameter space_npg2_adsds}.

\begin{figure}[!h]
  \begin{center}
    \includegraphics[width=.45\textwidth]{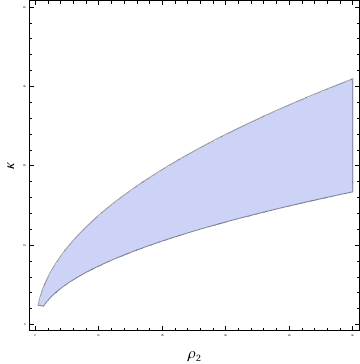}
  \end{center}
  \caption{Plot of the area in the parameter space spanned by $\rho_2$ and $\kappa$ for which no solutions to the field equations exist.}
  \label{fig:restriction on parameter space_mink}
\end{figure}

For compactifications on Minkowski backgrounds, the parameter $\mu_1$ vanishes.  The parameter spaces spanned by $\rho_2$ and $\kappa$ on the other hand is restricted by the same conditions as in the de Sitter and anti-de Sitter cases.  The area in the parameter spaces in which no solutions to the field equations with Minkowski background can be given is depicted in Figure \ref{fig:restriction on parameter space_mink}.

\subsection{\ghol compactifications}
\begin{figure}[!h]
  \begin{center}
    \begin{subfigure}[b]{.45\textwidth}
      \includegraphics[width=\textwidth]{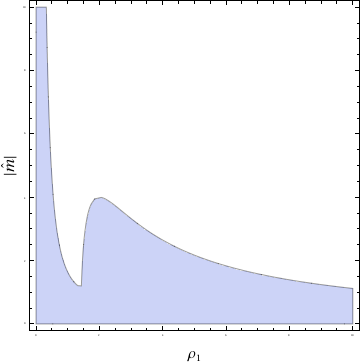}
      \label{fig:restriction on parameter space_ads}
      \caption{Anti-de Sitter backgrounds}
    \end{subfigure} 
    \begin{subfigure}[b]{.45\textwidth}
      \includegraphics[width=\textwidth]{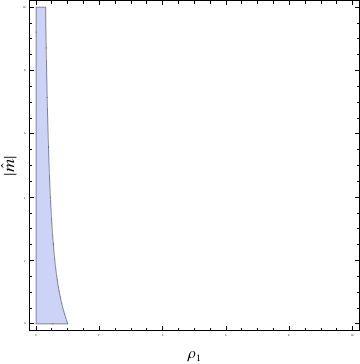}
      \label{fig:restriction on parameter space_ds}
      \caption{De Sitter backgrounds}
    \end{subfigure}
    \caption{Plot of the area in the parameter space spanned by $\rho_1$ and $\hat m$ for which no solutions to the field equations for \ghol compactifications exist.}
    \label{fig:restriction on parameter space_ghol}
  \end{center}
\end{figure}

The solutions to the equations for compactifications on \ghol manifolds as discussed in Section \ref{sec:hetStr_on_adsxX7} depend on the parameters $\rho_1$ and $\hat m$.  For anti-de Sitter backgrounds, as in the \np $G_2$ case, the field equations only possess solutions if the value $|\rho_1|$ exceeds a lower bound.  For de Sitter backgrounds on the other hand, there is no lower bound on $|\rho_1|$.  Additionally, on both backgrounds, there is a lower bound on $|\hat m|$.  Plot of the areas in the parameter space excluded by these lower bounds are shown in Figure \ref{fig:restriction on parameter space_ghol}.  Finally, for compactifications to Minkowski space on \ghol manifolds, the equations of motion are solvable for all values of the parameters.  

\end{appendix}

\newpage 

\bibliographystyle{JHEP}
\bibliography{library}

\providecommand{\href}[2]{#2}\begingroup\raggedright\begin{thebibliography}{10}

\bibitem{Strominger1986}
A.~Strominger, {\it Superstrings with torsion},  {\em Nuclear Physics B} {\bf
  274} (1986) 253.

\bibitem{Cardoso2003}
G.~Cardoso, G.~Curio, G.~Dall'Agata, D.~L\"{u}st, P.~Manousselis, and
  G.~Zoupanos, {\it {Non-Kähler string backgrounds and their five torsion
  classes}},  {\em Nuclear Physics B} {\bf 652} (2003) 5,
  [\href{http://xxx.lanl.gov/abs/hep-th/0211118}{{\tt hep-th/0211118}}].

\bibitem{Cardoso2004}
G.~L. Cardoso, G.~Curio, G.~Dall'Agata, and D.~L\"{u}st, {\it {Heterotic string
  theory on non-Kähler manifolds with H-flux and gaugino condensate}},  {\em
  Fortschritte der Physik} {\bf 52} (2004) 483,
  [\href{http://xxx.lanl.gov/abs/hep-th/0310021}{{\tt hep-th/0310021}}].

\bibitem{Frey2005}
A.~Frey and M.~Lippert, {\it {AdS strings with torsion: Noncomplex heterotic
  compactifications}},  {\em Physical Review D} {\bf 72} (2005) 42,
  [\href{http://xxx.lanl.gov/abs/hep-th/0507202}{{\tt hep-th/0507202}}].

\bibitem{Manousselis2006}
P.~Manousselis, N.~Prezas, and G.~Zoupanos, {\it Supersymmetric
  compactifications of heterotic strings with fluxes and condensates},  {\em
  Nuclear Physics B} {\bf 739} (2006) 85,
  [\href{http://xxx.lanl.gov/abs/hep-th/0511122}{{\tt hep-th/0511122}}].

\bibitem{Lechtenfeld2010}
O.~Lechtenfeld, C.~N\"{o}lle, and A.~D. Popov, {\it {Heterotic
  compactifications on nearly Kähler manifolds}},  {\em Journal of High Energy
  Physics} {\bf 1009} (2010) 74, [\href{http://xxx.lanl.gov/abs/1007.0236}{{\tt
  arXiv:1007.0236}}].

\bibitem{Chatzistavrakidis2012}
A.~Chatzistavrakidis, O.~Lechtenfeld, and A.~D. Popov, {\it {Nearly Kähler
  heterotic compactifications with fermion condensates}},  {\em Journal of High
  Energy Physics} {\bf 1204} (2012) 114,
  [\href{http://xxx.lanl.gov/abs/1202.1278}{{\tt arXiv:1202.1278}}].

\bibitem{Bergshoeff1982}
E.~Bergshoeff, M.~de~Roo, B.~de~Wit, and P.~van Nieuwenhuizen, {\it
  {Ten-dimensional Maxwell-Einstein supergravity, its currents, and the issue
  of its auxiliary fields}},  {\em Nuclear Physics B} {\bf 195} (1982) 97.

\bibitem{Chapline1983}
G.~Chapline and N.~Manton, {\it {Unification of Yang-Mills theory and
  supergravity in ten-dimensions}},  {\em Physics Letters B} {\bf 120} (1983)
  105.

\bibitem{Bergshoeff1989}
E.~A. Bergshoeff and M.~de~Roo, {\it The quartic effective action of the
  heterotic string and supersymmetry},  {\em Nuclear Physics B} {\bf 328}
  (1989) 439.

\bibitem{Fernandez1982}
M.~Fernández and A.~Gray, {\it {Riemannian manifolds with structure group
  $G_2$}},  {\em Annali di Matematica Pura ed Applicata} {\bf 132} (1982) 19.

\bibitem{Friedrich1997}
T.~Friedrich, I.~Kath, A.~Moroianu, and U.~Semmelmann, {\it {On nearly parallel
  $G_2$-structures}},  {\em Journal of Geometry and Physics} {\bf 23} (1997)
  259.

\bibitem{Bryant2006}
R.~L. Bryant, {\it {Some remarks on $G_2$-structures}},  in {\em Proceedings of
  G\"okova Geometry-Topology Conference 2005} (T.~O. S.~Akbulut and R.~Stern,
  eds.), pp.~75--109, International Press, 2006.

\bibitem{Bilal2001}
A.~Bilal, J.-P. Derendinger, and K.~Sfetsos, {\it {(Weak) $G_2$ holonomy from
  self-duality, flux and supersymmetry}},  {\em Nuclear Physics B} {\bf 628}
  (2002) 112, [\href{http://xxx.lanl.gov/abs/hep-th/0111274}{{\tt
  hep-th/0111274}}].

\end{thebibliography}\endgroup

\end{document}